\documentclass{ws-ijmpa}
\begin{document}

\markboth{Yu.~V.~Churkin  {\it et al.}}
{Dispersion Interaction of Atoms with Single-Walled
Carbon Nanotubes}

%
\catchline{}{}{}{}{}
%

\title{DISPERSION INTERACTION OF ATOMS WITH SINGLE-WALLED CARBON
NANOTUBES DESCRIBED BY THE DIRAC MODEL}

\author{
YU.~V.~CHURKIN, A.~B.~FEDORTSOV, G.~L.~KLIMCHITSKAYA and
V.~A.~YUROVA
}

\address{North-West Technical University,
 Millionnaya Street 5, St.Petersburg,
191065, Russia
}

\maketitle

\begin{history}
\received{2 June 2011}
\revised{1 July 2011}
\end{history}

\begin{abstract}
We calculate the interaction energy and force between atoms and
molecules and single-walled carbon nanotubes described by the
Dirac model of graphene. For this purpose the Lifshitz-type
formulas adapted for the case of cylindrical geometry with the help
of the proximity force approximation are used. The results obtained
are compared with those derived from the hydrodymanic model
of graphene.
Numerical computations are performed for hydrogen atoms and
molecules. It is shown that the Dirac model leads to larger
values of the van der Waals force than the hydrodynamic model.
For a hydrogen molecule the interaction energy and force computed
using both models are larger than for a hydrogen atom.

\end{abstract}

\ccode{PACS numbers: 12.20.-m, 42.50.Ct, 78.20.Ci}

\section{Introduction}

Dispersion interaction of atoms and molecules with different carbon
nanostructures is of topical interest for both fundamental and
applied physics. Carbon nanostructures possess unique electrical,
mechanical and optical properties\cite{1} leading to a variety of
unexpected phenomena. One of them is the possibility to absorb
hydrogen atoms and molecules that can be used for hydrogen
storage.\cite{2} Keeping in mind that for technological
purposes there is a need to create nanostructures capable of
absorbing more than 10 mass percent of hydrogen, various
theoretical methods allowing calculation of
atom-nanostructure interaction are of much current attention.

In the last few years dispersion interaction between atoms
(molecules) and graphene (a single sheet of graphite) and
single-walled carbon nanotubes was actively investigated
by using the phenomenological density-functional
theory.\cite{3}\cdash\cite{6}
The multiwalled carbon nanotubes with at least several
walls can be described by using the concept of the
dielectric permittivity of graphite. Because of this, the
dispersion interaction between atoms and multiwalled
nanotubes can be calculated\cite{7} using the Lifshitz
theory\cite{8} and the proximity force approximation
(PFA) (see Ref.~\refcite{9} for a review).
However, taking into account quick progress in
measurements of the van der Waals and Casimir
forces,\cite{10} the more fundamental calculation
methods of the dispersion interaction between atoms and
single-walled carbon nanotubes are also much needed.

Recent progress allowed generalization of the Lifshitz
theory for the material bodies of arbitrary shape with
known reflection properties.\cite{9,11}\cdash\cite{13}
Along these lines the dispersion interaction between
carbon nanostructures and material bodies can be
described if the reflection properties of the electromagnetic
oscillations on graphene are available.
In the framework of the so-called {\it hydrodynamic}
model, graphene was considered as an infinitesimally thin
positively charged flat sheet, carring a homogeneous
fluid with some mass and negative charge densities.\cite{14}
The reflection coefficients of electromagnetic oscillations
on such a sheet were found in Refs.~\refcite{15} and
\refcite{16}. Using these reflection coefficients, the
Lifshitz theory was applied to calculate the dispersion
interaction between two parallel graphene sheets,\cite{17}
between graphene and a material plate,\cite{18} and
between graphene and an atom or a molecule.\cite{19}
By combining the original formulation of the Lifshitz
theory (which is applicable to only plane parallel
systems) with the PFA, it was possible also to
calculate the dispersion  interaction between
an atom or a molecule and a single-walled carbon
nanotube\cite{19} in the framework of the hydrodynamic
model of graphene.

A more exact description of graphene is given by the
{\it Dirac} model\cite{20}\cdash\cite{22} which takes
into account that the quasiparticle fermion excitations in
graphene are the massless Dirac fermions which move with
a Fermi velocity and have a linear dispersion relation.
This description is also an approximation (it is valid
at low energies up to a few eV). Under the assumption
that the Dirac model holds at any energy, the reflection
coefficients for the electromagnetic oscillations on
graphene were found in Ref.~\refcite{23}.
The Lifshitz theory combined with the reflection
coefficients of the Dirac model has been used to calculate
the Casimir interaction between a graphene and a parallel
ideal metal plane\cite{23} and between atoms and molecules
and graphene.\cite{24}
The computational results obtained using the hydrodynamic
and Dirac models were compared.\cite{24}
Specifically, an experiment on quantum reflection was
proposed\cite{24} allowing to discriminate between the
predictions of the hydrodynamic and Dirac models of
graphene. The influence of the thermal effects in the
framework of the Dirac model was investigated in
Ref.~\refcite{25}.

In this paper, we calculate the energy and force of
dispersion interaction between hydrogen atoms and
molecules and single-walled carbon nanotubes using the
Dirac model. The obtained results are compared with
those obtained earlier using the hydrodynamic model.
Keeping in mind that at short separations from
1 to 3\,nm, which are most interesting for the
investigation of absorption processes, the thermal
effects are negligibly small, we consider the Dirac
model at zero temperature. We arrive at the conclusion
that the van der Waals coefficient of the interaction
energy computed using the Dirac model can be both
larger and smaller than the same coefficient
computed using the hydrodynamic model depending on
separation. As to the coefficient of the van der Waals
force, it is always larger when the Dirac model is
used in computations.

The paper is organized as follows. In Sec.~2 we present
the Lifshitz-type formulas for the atom-nanotube
interaction and the reflection coefficients for both
models of graphene. Section~3 contains the results of
numerical computations of the van der Waals energy and
force. In Sec.~4 the reader will find our conclusions
and discussion.

\section{Energy and Force of Dispersion Interaction
Between an Atom and a Nanotube}

We use the Lifshitz-type formulas for the van der Waals
and Casimir-Polder interaction of atoms (molecules)
with single-walled carbon nanotubes obtained using the
PFA.\cite{7,9,19} The validity of the PFA for the
configuration of an atom near an ideal-metal
cylindrical shell, where the exact expression for the
Casimir-Polder potential is available, was recently
demonstrated in Ref.~\refcite{26}. Specifically, it
was shown\cite{26} that for $a/R=0.1$ ($R$ is the
cylinder radius and $a$ is the separation distance
between an atom and a cylindrical surface) the relative
deviation between the exact and the PFA results is of
about 1\%. At $a/R=0.6$ this deviation does not exceed
4\%. The same measure of agreement between the exact
and the PFA results for atom-cylinder interaction is
expected for cylinders made of real materials.

The Lifshitz-type formulas for the interaction energy
and force between an atom (molecule) and a single-walled
nanotube in the framework of the PFA are given by
\begin{eqnarray}
&&
E(a)=-\frac{\hbar}{2\pi}\sqrt{\frac{R}{R+a}}
\int_{0}^{\infty}d\xi\,\alpha(i\xi)
\int_{0}^{\infty}k\,dk\,e^{-2aq}
\nonumber \\[1mm]
&&~~~~~~~\times
\left[q-\frac{1}{4(R+a)}\right]\,\left[2r_{\rm TM}-
\frac{\xi^2}{q^2c^2}(r_{\rm TM}+r_{\rm TE})\right],
\label{eq1} \\[2mm]
&&
F(a)=-\frac{\hbar}{2\pi}\sqrt{\frac{R}{R+a}}
\int_{0}^{\infty}d\xi\,\alpha(i\xi)
\int_{0}^{\infty}k\,dk\,e^{-2aq}
\nonumber \\[1mm]
&&~~~~~~~\times
\left[2q^2-\frac{3}{8(R+a)^2}\right]\,\left[2r_{\rm TM}-
\frac{\xi^2}{q^2c^2}(r_{\rm TM}+r_{\rm TE})\right].
\nonumber
\end{eqnarray}
\noindent
Here, $\alpha(\omega)$ is the dynamic polarizability of
an atom or a molecule, $\omega=i\xi$ is the imaginary
frequency, $q^2=k^2+\xi^2/c^2$, and $r_{\rm TM}$ and
$r_{\rm TE}$ are the reflection coefficients of the
electromagnetic fluctuations on graphene for two
independent polarizations of the electromagnetic field,
transverse magnetic and transverse electric.

The explicit expressions for the reflection coefficients
depend on the model of the electronic structure of
graphene used. In the framework of the Dirac model
these coefficients can be presented in the form\cite{23}
\begin{eqnarray}
&&
r_{\rm TM}\equiv r_{\rm TM}^{(D)}(i\xi,k)
=\frac{\alpha q\Phi(\tilde{q})}{2\tilde{q}^2+\alpha q\Phi(\tilde{q})},
\nonumber \\[1mm]
&&
r_{\rm TE}\equiv r_{\rm TE}^{(D)}(i\xi,k)
=-\frac{\alpha\Phi(\tilde{q})}{2q+\alpha\Phi(\tilde{q})},
\label{eq2}
\end{eqnarray}
\noindent
where $\alpha=e^2/(\hbar c)\approx1/137$ is the fine-structure constant,
$\tilde{q}^2=(v_{\rm F}^2k^2+\xi^2)/c^2$,
$v_{\rm F}\approx 10^{6}\,$m/s is the Fermi velocity, and the function
$\Phi$ determines the polarization tensor in an external electromagnetic
field in the one-loop approximation in three dimensional space-time.
The explicit form of this function along the imaginary frequency axis
is the following:\cite{23}
\begin{equation}
\Phi(\tilde{q})=N\left(\tilde\Delta+
\frac{\tilde{q}^2-4\tilde\Delta^2}{2\tilde{q}}\,
\arctan\frac{\tilde{q}}{2\tilde\Delta}\right).
\label{eq3}
\end{equation}
\noindent
Here, $N=4$, $\tilde\Delta=\Delta/(\hbar c)$, and the exact
value of the gap parameter $\Delta$ remains unknown. The upper bound
on $\Delta$ is equal to about 0.1\,eV, but the true value of
$\Delta$ might be much smaller.\cite{22}

In the framework of the hydrodynamic model the reflection coefficients
are presented differently
\begin{eqnarray}
&&
r_{\rm TM}\equiv r_{\rm TM}^{(h)}(i\xi,k)
=\frac{c^2qK}{c^2qK+\xi^2},
\nonumber \\[1mm]
&&
r_{\rm TE}\equiv r_{\rm TE}^{(h)}(i\xi,k)
=-\frac{K}{K+q},
\label{eq4}
\end{eqnarray}
\noindent
where the wave number of the graphene sheet
$K=6.75\times 10^{5}\,\mbox{m}^{-1}$ corresponds to the frequency
$\omega_{K}=cK=2.02\times 10^{14}\,$rad/s.

For the purposes of numerical computations it is convenient to
introduce the separation-dependent van der Waals coefficient
$C_3(a)$ and the force coefficient $C_F(a)$, so that the
interaction energy and force (\ref{eq1}) are represented as
\begin{equation}
E(a)=-\frac{C_3(a)}{a^3}, \qquad
F(a)=-\frac{C_F(a)}{a^4}.
\label{eq5}
\end{equation}
\noindent
The explicit form of the coefficients $C_3(a)$ and  $C_F(a)$
in terms of  dimensionless variable $y=2aq$ is obtained
from Eq.~(\ref{eq1})
\begin{eqnarray}
&&
C_3(a)=\frac{\hbar}{16\pi}\sqrt{\frac{R}{R+a}}
\int_{0}^{\infty}d\xi\,\alpha(i\xi)
\int_{2a\xi/c}^{\infty}y\,dy\,e^{-y}
\nonumber \\[1mm]
&&~~~~~~~\times
\left[y-\frac{a}{2(R+a)}\right]\,\left[2r_{\rm TM}-
\frac{4a^2\xi^2}{y^2c^2}(r_{\rm TM}+r_{\rm TE})\right],
\label{eq6} \\[2mm]
&&
C_F(a)=\frac{\hbar}{16\pi}\sqrt{\frac{R}{R+a}}
\int_{0}^{\infty}d\xi\,\alpha(i\xi)
\int_{2a\xi/c}^{\infty}y\,dy\,e^{-y}
\nonumber \\[1mm]
&&~~~~~~~\times
\left[y^2-\frac{3a^2}{4(R+a)^2}\right]\,\left[2r_{\rm TM}-
\frac{4a^2\xi^2}{y^2c^2}(r_{\rm TM}+r_{\rm TE})\right].
\nonumber
\end{eqnarray}
\noindent
Equations (\ref{eq5}) and (\ref{eq6}) can be used with either
the Dirac- or hydrodynamic-model reflection coefficients
taking into account that in terms of the variable $y$ it holds
\begin{equation}
q=\frac{y}{2a},\quad
\tilde{q}=\left[\frac{v_{\rm F}^2}{c^2}\frac{y^2}{4a^2}+
\left(1-\frac{v_{\rm F}^2}{c^2}\right)\frac{\xi^2}{c^2}
\right]^{1/2}.
\label{eq7}
\end{equation}
\noindent
For an atom interacting with a graphene sheet the interaction
energy also has the form of Eq.~(\ref{eq5}), but the
van der Waals coefficient is given by\cite{24}
\begin{eqnarray}
&&
C_3(a)=\frac{\hbar}{16\pi}
\int_{0}^{\infty}d\xi\,\alpha(i\xi)
\int_{2a\xi/c}^{\infty}dy\,e^{-y}
\nonumber \\[1mm]
&&~~~~~~~\times
\left[2y^2r_{\rm TM}-
\frac{4a^2\xi^2}{c^2}(r_{\rm TM}+r_{\rm TE})\right].
\label{eq8}
\end{eqnarray}

Note that if the coefficient $C_3$ would not depend on $a$ and
the equality $C_F=3C_3$ be satisfied,
then Eq.~(\ref{eq5})describes the nonretarded van der Waals
interaction between an atom and a material
wall.\cite{8}\cdash\cite{10,27}\cdash\cite{30}
Numerical computations performed in the next section show that
for hydrogen atoms and molecules interacting with a graphene
sheet and single-walled carbon nanotubes the coefficients
$C_3$ and $C_F$ are separation-dependent down to the shortest
separation distances of about 1\,nm where a macroscopic
description of dispersion interaction by means of the
Lifshitz-type formulas remains applicable.

\section{Dispersion Interaction of Hydrogen Atoms and
Molecules with Single-Walled Carbon Nanotubes}

In this section, we perform numerical computations of the
coefficients $C_3$ and $C_F$ in Eq.~(\ref{eq6})
describing the energy and force of the van der Waals
interaction, respectively. All computations are performed
for hydrogen atoms and molecules using the Dirac and
hydrodynamic models of graphene, and the obtained results
are compared. The atomic and molecular dynamic
polarizabilities of hydrogen can be represented with
sufficient precision using the single-oscillator
model.\cite{7,30} According to this model, the dynamic
polarizabilities along the imaginary frequency axis take
the form
\begin{equation}
\alpha(i\xi)=\frac{\alpha(0)}{1+\frac{\xi^2}{\omega_0^2}},
\label{eq8a}
\end{equation}
\noindent
where $\alpha(0)$ is the static atomic (molecular)
polarizability and $\omega_0$ is the
characteristic frequency. For a hydrogen atom it was found\cite{31}
that $\alpha(0)\equiv\alpha_a(0)=4.50\,$a.u. (1\,a.u.
of polarizability is equal to
$1.482\times 10^{-31}\,\mbox{m}^{3}$)
and $\omega_0\equiv\omega_{0a}=11.65\,$eV.
For a hydrogen molecule\cite{31}
$\alpha(0)\equiv\alpha_m(0)=5.439\,$a.u. and
$\omega_0\equiv\omega_{0m}=14.09\,$eV.

\begin{figure*}[t]
\vspace*{-2.7cm}
\centerline{\hspace*{2.1cm}\psfig{file=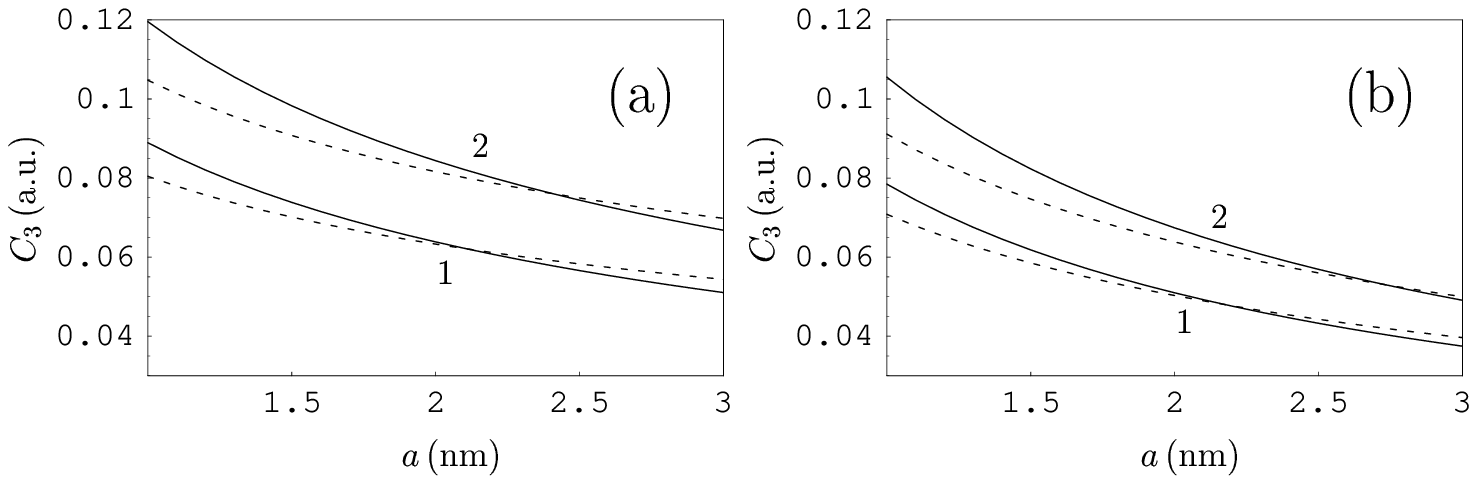,width=17.6cm}}
\vspace*{-17.5cm}
\caption{
The van der Waals coefficients for hydrogen atom (the solid and
dashed lines labeled 1) and molecule (the solid and
dashed lines labeled 2) interacting with (a) a graphene sheet
and (b) a single-walled carbon nanotube of $R=5\,$nm radius as
a function of separation. Computations are performed using the
Dirac (the solid lines) and hydrodynamic (the dashed lines)
models of graphene.
}
\end{figure*}
In Fig.~1(a) we present the computational results for the van der Waals
coefficient $C_{3,\rm H}$ in atomic units as a function of separation
for the interaction of a hydrogen atom  (the dashed and solid lines labeled 1)
and molecule (the dashed and solid lines labeled 2) with graphene.
One atomic unit for $C_3$ is equal to
$4.032\times 10^{-3}\,\mbox{eV\,nm}^{3}$.
The solid lines 1 and 2 are computed using the Dirac model (\ref{eq2})
with the gap parameter $\Delta=0.1\,$eV (as shown in Ref.~\refcite{24},
the computational results only slightly depend on the value of
$\Delta$). The dashed lines 1 and 2 are computed using the
hydrodynamic model (\ref{eq4}).
Figure 1(a) extends the results of Ref.~\refcite{24} for
separations below 3\,nm. At $a>3\,$nm the van der Waals coefficient
$C_3^{(h)}$ computed using the hydrodynamic model is always larger
than $C_3^{(D)}$ computed using the Dirac model.\cite{24}
As can be seen in Fig.~1(a), at $a<3\,$nm this is, however, not so.
Thus, for hydrogen atom at $a<2.1\,$nm $C_3^{(h)}<C_3^{(D)}$.
For hydrogen molecule the same holds at $a<2.42\,$nm.

In Fig.~1(b) similar results are shown for hydrogen atom and
molecule interacting with the single-walled carbon nanotube
of 5\,nm radius. All notations are the same as in Fig.~1(a).
The maximum separation distance $a=3\,$nm is chosen in order to
preserve the parameter $a/R\leq 0.6$, i.e., to avoid large
deviations from the PFA. As is seen in Fig.~1(b), for
single-walled carbon nanotubes there are also some small
separation distances ($a=2.2\,$nm for a hydrogen atom and
$a=2.77\,$nm for a hydrogen molecule) where the values of the van
der Waals coefficients computed using the Dirac and
hydrodynamic model become equal.

Now we compute the coefficient of the van der Waals force $C_F$
in Eq.~(\ref{eq6}) for an atom and a molecule interacting with a
single-walled carbon nanotube. The computational results as a function
of separation are presented in Fig.~2(a) for a hydrogen atom and
in Fig.~2(b) for a hydrogen molecule. In both cases the nanotube
radius is $R=5\,$nm. As can be seen in Fig.~2(a,b), independently
of the model of graphene used (the solid lines are for the Dirac
model with the gap parameter $\Delta=0.1\,$eV and the dashed
lines are for the hydrodynamic model), $C_F$ is the monotonously
decreasing function with the increase of separation.
The largest relative differences between the values of $C_F$
computed using the Dirac and the hydrodynamic model (29\% for
a hydrogen atom and 30\% for a hydrogen molecule) are
achieved at $a=1\,$nm.
\begin{figure*}[t]
\vspace*{-2.7cm}
\centerline{\hspace*{2.1cm}\psfig{file=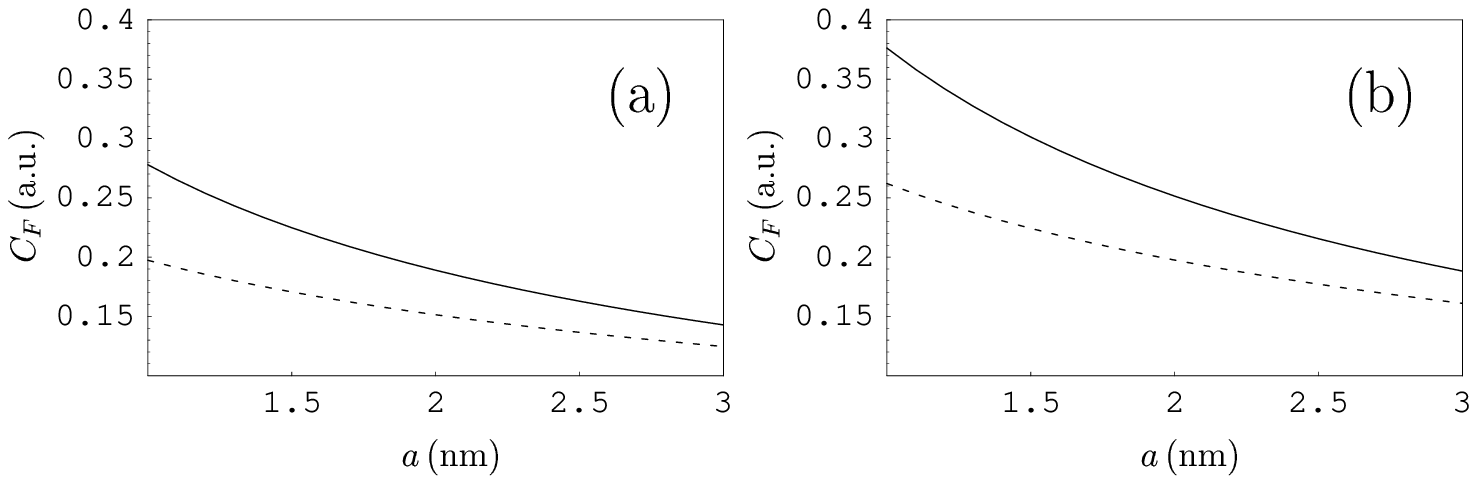,width=17.6cm}}
\vspace*{-17.5cm}
\caption{
The coefficients of the van der Waals  force
for hydrogen (a) atom  and (b) molecule
interacting with  a single-walled carbon nanotube of $R=5\,$nm radius as
a function of separation. Computations are performed using the
Dirac (the solid lines) and hydrodynamic (the dashed lines)
models of graphene.
}
\end{figure*}

\begin{figure*}[b]
\vspace*{-2.7cm}
\centerline{\hspace*{2.1cm}\psfig{file=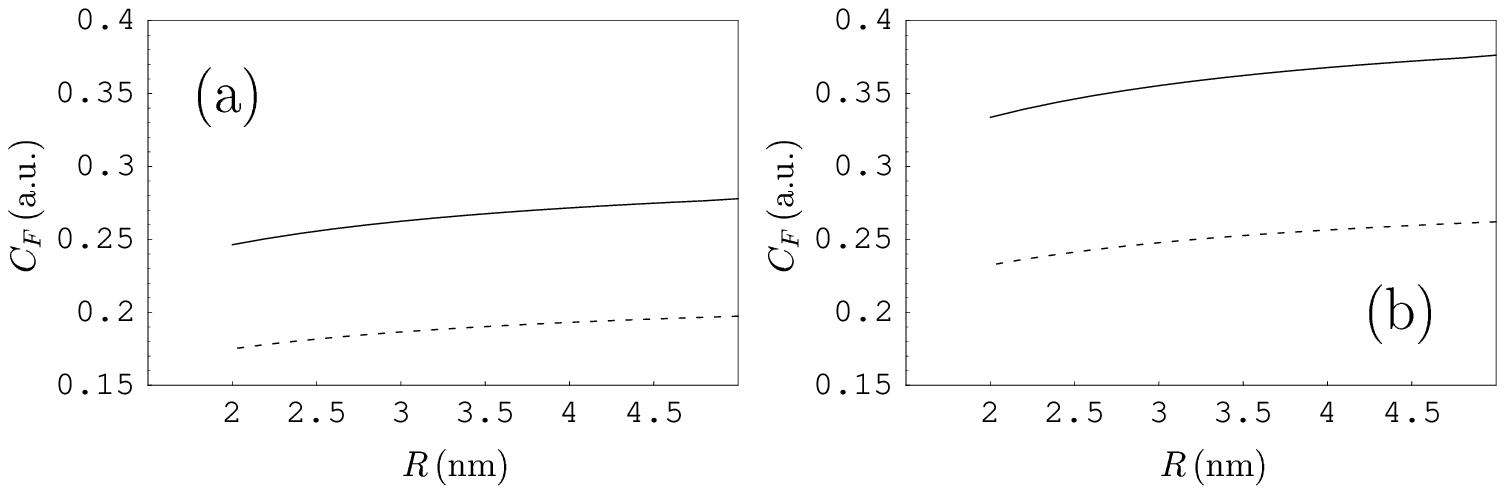,width=17.6cm}}
\vspace*{-17.5cm}
\caption{
The coefficients of the van der Waals  force
for hydrogen (a) atom  and (b) molecule
interacting with  a single-walled carbon nanotube as
a function of its radius. Computations are performed using the
Dirac (the solid lines) and hydrodynamic (the dashed lines)
models of graphene for an atom or a molecule at the
separation $a=1\,$nm from the nanotube surface.
}
\end{figure*}
Further we investigate the dependence of the coefficient $C_F$ in
Eq.~(\ref{eq6}) on the radius of a single-walled carbon nanotube
$R$. In Fig.~3(a) we plot the coefficient $C_F$ as a function
of $R$ for the dispersion interaction of a hydrogen atom
separated by the distance $a=1\,$nm from the surface of a
carbon nanotube. The solid line is obtained by using the
reflection coefficients (\ref{eq2}) of the Dirac model while
the dashed line is computed with the help of reflection
coefficients (\ref{eq4}) of the hydrodynamic model.
As can be seen in Fig.~3(a), in both models the coefficient
$C_F$ slowly increases with increasing radius of the nanotube.
For all radii, the value of $C_F$ computed within the Dirac
model is larger than within the hydrodynamic model.
For example, for a nanotube with $R=2\,$nm the relative
difference between the predictions of the Dirac and
hydrodynamic models is equal to approximately 30\%.
Thus, it is several times larger than the error
arising from using
the PFA (see Sec.~2). In Fig.~3(b) the coefficient $C_F$
as a function of nanotube radius is plotted for a
hydrogen molecule at the separation $a=1\,$nm from a
single-walled carbon nanotube. The same notations, as in
Fig.~3(a), are used. Here, for a nanotube of $R=2\,$nm
radius the relative
difference between the predictions of the Dirac and
hydrodynamic models is equal to 29\%.
{}From Fig.~3 it follows that there is only a minor
dependence of the relative difference in the predictions
of the Dirac and hydrodynamic models for the coefficients
$C_3$ and $C_F$ on a nanotube radius.

\section{Conclusions and Discussion}

In the foregoing we have investigated the dispersion
interaction of hydrogen atoms and molecules with
single-walled carbon nanotubes described by the Dirac and
hydrodynamic models. The theoretical basis for our study
is the Lifshitz theory expressing the interaction energy
and force between planar material structures in terms
of the reflection coefficients of electromagnetic
fluctuations. Using the PFA, the Lifshitz formulas
for the van der Waals energy and force were adapted
for the configuration of an atom or a molecule
interacting with a nanotube of cylindrical shape.

Computations of the van der Waals coefficient $C_3$
performed in the paper demonstrated that its
magnitude for a hydrogen molecule is always significantly
larger than for a hydrogen atom.
At the same time, the value of $C_3$ for an atom-graphene
(molecule-graphene) interaction is only slightly larger
than for an atom-nanotube (molecule-nanotube)
interaction. Different models of graphene (Dirac and
hydrodynamic) lead to a bit different predictions
for $C_3$ as a function of separation.
Furthermore, there is the separation distance below
3\,nm where both models result in one and the same
value of $C_3$.

Similar results follow from computations of the
coefficient $C_F$ characterizing the van der Waals
force. Its magnitude for a hydrogen molecule is
also significantly larger than for a hydrogen atom.
It is interesting that the Dirac model
of graphene leads to markedly larger results
for  $C_F$ than the
hydrodynamic model. In the application region of the
PFA, there is no separation distance where the values
of $C_F$ predicted by the two models would be equal.
We demonstrated that the above results are preserved for
single-walled carbon nanotubes of different radii
from 2 to 5\,nm. Keeping in mind that the Dirac model
gives a more accurate description of graphene at low
energies than the hydrodynamic model, the former can be
applied to calculate the capacity of single-walled
carbon nanotubes to absorb hydrogen atoms and
molecules for use in hydrogen storage.

\section*{Acknowledgments}

This work  was supported by the Grant of the Russian Ministry of
Education P--184.


\end{document}